\documentclass[%
 aip,
 amsmath,amssymb,
 reprint,
]{revtex4-1}

\usepackage{graphicx}% Include figure files
\usepackage{dcolumn}% Align table columns on decimal point
\usepackage{bm}% bold math
\usepackage[utf8]{inputenc}
\usepackage[T1]{fontenc}
\usepackage{mathptmx}
\DeclareMathAlphabet{\mathcal}{OMS}{cmsy}{m}{n}
\usepackage{etoolbox}

\makeatletter
\def\@email#1#2{%
 \endgroup
 \patchcmd{\titleblock@produce}
  {\frontmatter@RRAPformat}
  {\frontmatter@RRAPformat{\produce@RRAP{*#1\href{mailto:#2}{#2}}}\frontmatter@RRAPformat}
  {}{}
}%
\makeatother

\begin{document}

\title[High efficiency superconducting filterbank with impedance-defined resolution for millimeter-wave spectroscopy]{High efficiency superconducting filterbank with impedance-defined resolution for millimeter-wave spectroscopy}

\author{O. Jeong}
\email{objeong@lbl.gov}
\affiliation{CNRS-UCB International Research Laboratory, Centre Pierre Bin\'etruy, \\IRL 2007, CPB-IN2P3, Berkeley, CA 94720, USA}
\affiliation{Lawrence Berkeley National Laboratory, 1 Cyclotron Road, Berkeley, CA 94720, USA}
\author{M. Piat}%
\affiliation{Universit\'e Paris Cit\'e, CNRS, Astroparticule et Cosmologie, F-75013 Paris, France}

\author{A. Suzuki}
\affiliation{Lawrence Berkeley National Laboratory, 1 Cyclotron Road, Berkeley, CA 94720, USA}

\date{\today}

\begin{abstract}
We present a high efficiency, moderate resolution on-chip superconducting filterbank spectrometer designed for line intensity mapping and broadband wave-like dark matter searches. Existing implementations used by the millimeter-wave community rely on resistive feedline termination for standing wave mitigation which caps the per-channel efficiency $\eta_\mathrm{ch}$ at 50\% and utilize microstrip resonators which highly couple the resolving power $\mathcal{R}=\nu/\delta\nu$ and $\eta_\mathrm{ch}$ to thin-film dielectric loss. The architecture presented in this paper addresses these limitations by eliminating the termination resistor and employing niobium coplanar waveguide resonators patterned directly on single-crystal silicon, which fixes the resonator's internal quality factor by the silicon substrate property rather than by the loss tangent of a deposited film. The resolving power is set by the impedance ratio between the resonator and feedline. Electromagnetic and circuit simulations of a ten-channel filterbank near 90~GHz at unity oversampling show $\mathcal{R}=1170$ and $\eta_\mathrm{ch}> 78\%$, with no significant trend in efficiency across channels. A ten-channel prototype, designed for $\mathcal{R}=92\pm5$ and $\eta_\mathrm{ch}=(81\pm3)\%$ over 90--106~GHz with a 560~nm interlayer based on simulations, was fabricated and optically characterized, yielding $\mathcal{R}=208\pm74$ and $\eta_\mathrm{ch}=(77\pm6)\%$. Measured $\eta_\mathrm{ch}$ well above the 50\% termination limit, together with demonstrated impedance tuning of $\mathcal{R}$, validate the design principles of this filterbank spectrometer and establishes a route to high-efficiency, high-resolution on-chip spectrometer arrays.
\end{abstract}

\maketitle

\section{\label{sec:intro}Introduction}

Millimeter-wave spectroscopy provides a unique window into fundamental cosmology and particle physics. In line intensity mapping (LIM), medium resolution spectrometer translates measured atomic and molecular emission frequencies into precise line-of-sight redshifts. LIM offers an efficient pathway to probe the high redshift universe ($z\gtrsim 4$), extending the success of optical surveys like the Dark Energy Spectrscopic Instrument.~\cite{desi_Adame_2025} This enables three-dimensional mapping of cosmic large-scale structure across unprecedented volumes, unlocking new opportunities to study fundamental topics in cosmology such as multi-field inflation and dynamic dark energy.  Similarly, in broadband wave-like dark matter (DM) search experiments such as BREAD,~\cite{bread} a spectrometer can partition broadband signals into narrow frequency bins that map directly to discrete DM mass intervals, providing high precision particle mass identification.

Several spectrometer technologies have been developed, including Fourier transform spectrometers,~\cite{fts} diffraction gratings,~\cite{grating_time} on-chip diffraction gratings,~\cite{microspec_2014} and on-chip filterbank spectrometers.~\cite{SuperSpec_redford_2022, DESHIMA_Endo_2019, DESHIMA2, waveguide_nie_2024, waveguide_2025, marting_2026} Among these, on-chip filterbanks are particularly promising for LIM due to their inherent scalability required for deep surveys.~\cite{snowmass} Their compact physical footprint allows for large pixel count focal planes and their ability to couple hundreds of narrow-band spectroscopic channels to a single feedline enables probing a large range of frequencies, and thus redshifts, simultaneously with a single pixel. While technologies such as SuperSpec~\cite{SuperSpec_redford_2022} and DESHIMA~\cite{DESHIMA_Endo_2019} have demonstrated the viability of this design approach, existing implementations are limited in optical efficiency ($\eta_\mathrm{ch}$, the fraction of power entering the feedline that reaches a given channel's detector) and resolving power ($\mathcal{R} = \nu/\delta \nu$) due to both fundamental circuit design constraints and by the loss properties of the deposited amorphous dielectric film with which their resonators are built.

In this paper, we present a new filterbank architecture that simultaneously delivers high optical efficiency and medium resolution. We describe the governing circuit principles, verify performance via electromagnetic simulation, and experimentally demonstrate the architecture using a prototype device that achieves $\mathcal{R} = 208 \pm 74$ and $\eta_\mathrm{ch} = (77 \pm 6)\%$, successfully validating our design goals.

\section{Design Principles\label{sec:principles}}

\begin{figure}[htbp]
\includegraphics[width=0.75\columnwidth]{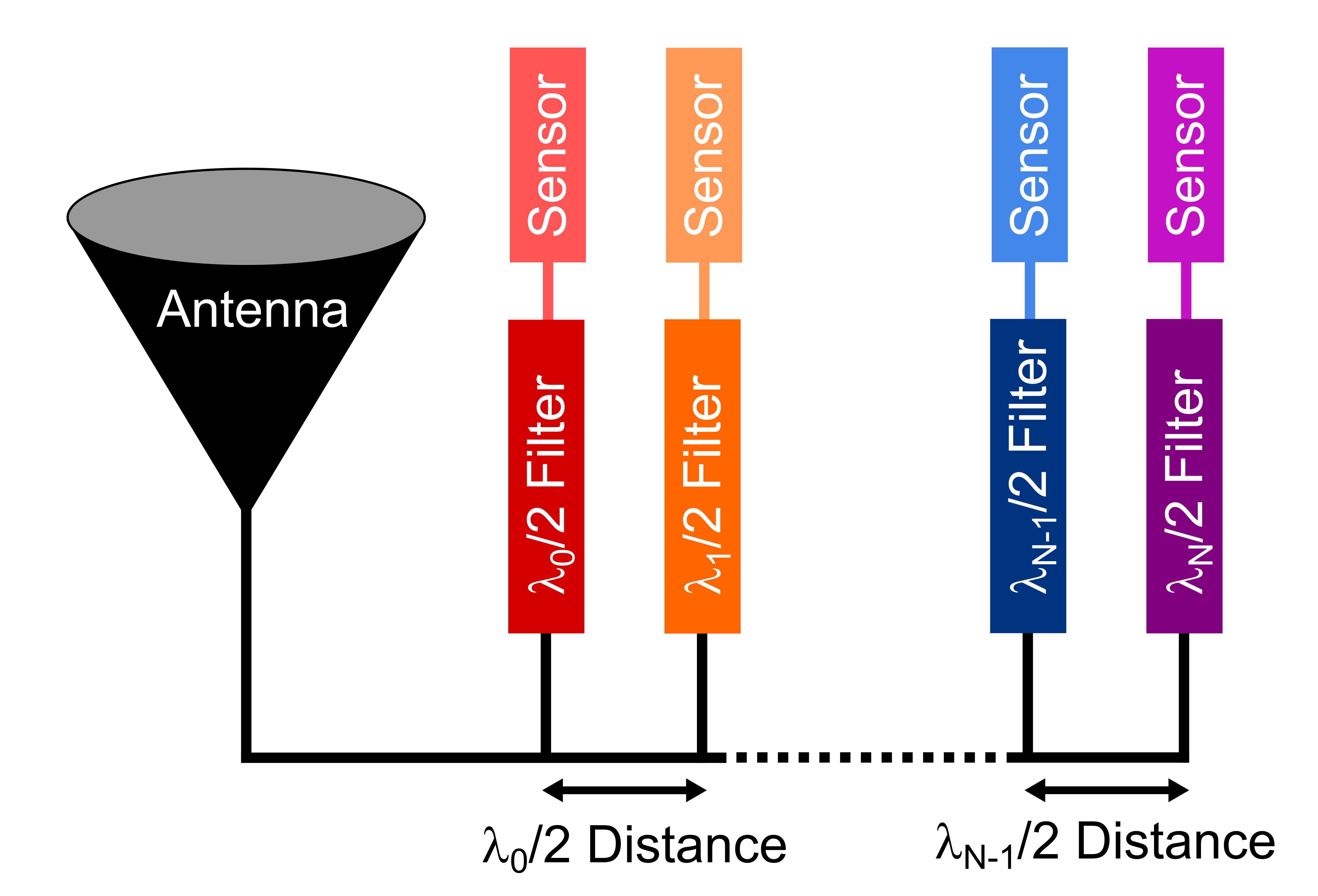}
\caption{\label{fig:toydesign} Design schematic of the N-channel filterbank spectrometer. Broadband signal enters from the antenna and propagates along a feedline tapped by half-wave resonant filters, each terminated on its own sensor; adjacent taps are separated by half-wavelength distance of the prior channel and the feedline is left open at its far end.}
\end{figure}

This spectrometer implements a cochlear channelizer, where power propagates along a common transmission line coupled to a sequence of log-periodically spaced resonant bandpass filters (see Fig.~\ref{fig:toydesign}),~\cite{obrient1, obrient2, galbraith, rebeiz} like the frequency-selective mechanics of the human ear. In this implementation, signal from the sky is captured by an antenna and coupled into the main transmission feedline. As the signal propagates along the feedline, it encounters a sequence of shunts arranged in descending order of frequency. Each shunt leads to a distributed half-wave resonant structure with a common resolving power. These resonators are designed to selectively couple power within a narrow bandwidth at unique, staggered frequencies, channelizing the incoming broadband signal for medium resolution spectroscopy. The center frequency $\nu_i$ of such half-wave resonator is fixed by its physical length $l_i$ following $\nu_i \propto 1/l_i$. Tiling a band at constant $\mathcal{R}$ therefore requires resonator lengths in geometric progression,
\begin{align}
\label{eq:step_factor}
    &l_i = l_0 \cdot r^i \\
    &r = 1 +\frac{1}{\mathcal{R}\Sigma}, 
\end{align}
where $\Sigma$ is the oversampling factor which sets the channel spacing relative to the channel width. Written as a sum of inverse quality factors, the loaded quality factor of an isolated channel defines the baseline spectral resolution $\mathcal{R} = Q$,

\begin{equation}
\label{eq:R}
   \frac{1}{\mathcal{R}}=\frac{1}{Q}=\frac{1}{Q_\mathrm{feed}}+\frac{1}{Q_\mathrm{det}}+ \frac{1}{Q_\mathrm{int}},
\end{equation}
where $Q_\mathrm{feed}$ is the coupling quality factor between the feedline and filter, $Q_\mathrm{det}$ is the coupling quality factor between the filter and detector, and $Q_\mathrm{int}$ is the internal quality factor describing dissipation within the resonator.

Existing millimeter-wave spectrometers suffer from an efficiency-resolution bottleneck driven by two primary circuit design elements. First, conventional on-chip filterbanks employ a resistive terminator after the final channel to prevent standing waves and residual reflections;~\cite{Barry} however, this design choice caps the channel efficiency at $\eta_\mathrm{ch} = 50\%$ before accounting for any material dissipation.~\cite{hd_superspec} Here $\eta_\mathrm{ch}$ denotes the fraction of the power incident at the feedline input that is delivered to a given channel's detector at its resonant frequency, the convention used throughout this paper. Efficiency referenced to the feedline input rather than to the power arriving at an individual shunt is the conservative convention, and the two differ increasingly for channels far along a lossy bank; conventions vary in the literature and quoted figures should be compared with this in mind. The highest efficiency measured in lab from a fabricated device with $\Sigma\geq1$ is 16\%.~\cite{DESHIMA2} Note that spectrometers can oversample across the band to boost the cumulative efficiency of the spectrometer. However, covering the same frequency range at $\Sigma\geq1$ requires proportionally more detectors and beyond Nyquist sampling ($\Sigma=2$), there are diminishing gains as one pays the penalty in detector noise for splitting signal at a single frequency between more and more detectors.~\cite{Wheeler} This motivates the development of a filterbank technology with $\eta_\mathrm{ch} > 50\%$ as it ensures manageable detector counts and optimal noise performance. 

Second, existing on-chip implementations rely on half-wave microstrip (MS) resonators~\cite{SuperSpec_redford_2022, spt_slim_karkare, DESHIMA2} which utilize lossy amorphous dielectric interlayers. This makes the resolving power $\mathcal{R}$ and efficiency $\eta_\mathrm{ch}$ strongly sensitive to the material's loss tangent ($\tan \delta$), posing a significant risk of performance degradation during fabrication. For instance, recent implementations designed around nominal silicon nitride (SiNx) loss tangent ($\tan\delta\sim10^{-3}$) suffered significant degradation of resolution when fabrication variation led to unexpectedly high dielectric loss.~\cite{Pan_2023, SiN_verylow, spt_slim} While extensive efforts across the community have targeted lower-loss alternatives, such as hydrogenated amorphous silicon ($\text{a-Si:H}$)~\cite{aSiH} or single-crystal silicon-on-insulator (SOI) layers,~\cite{class_SOI} these approaches entail demanding material process development. The spectrometer presented in this paper addresses both problems with two targeted modifications.

The first modification is removal of the terminating resistor. In the terminated architecture used by existing on-chip filterbanks,~\cite{Barry, SuperSpec_redford_2022, DESHIMA2} the resistor absorbs power which survives to the end of the feedline: out-of-band power that no channel can accept, and the in-band residue left by imperfect coupling at each shunt. Removing the resistor returns that power to the circuit. Uncaptured signal reflects from the open end of the feedline and coherently interferes with the incoming wave, so power missed on one pass can be absorbed on a later one. This is implemented by spacing the shunts at local half-wave distances: channel $i$ and $i+1$ are placed at $\lambda_i/2$ distance apart. Per-channel efficiency is thereby kept near 100\% rather than capped at 50\% (see supplementary material Sec.~I). Note that Marting \textit{et al.}~\cite{marting_2026} have recently demonstrated that directional filters, likewise dispensing of the feedline termination, can push per-channel efficiency well past 50\%, although achieving this together with high $\mathcal{R}$ requires further development.

The second modification is a coplanar waveguide (CPW) resonator with niobium (Nb) patterned directly on single-crystal silicon (Si). This confines the resonator fields to a substrate where the dielectric constant and loss tangent are well-controlled, with the loss tangent ($\tan\delta\le 10^{-5}$) being at least an order of magnitude lower than that of deposited amorphous films.~\cite{Si_loss,Datta_2013} As a result, $Q_\mathrm{int}$ far exceeds the coupling $Q$-factors such that $\mathcal{R}$ in Eq.~\ref{eq:Q_impedance} is set by geometry rather than by material. An MS resonator on a crystalline interlayer or a-Si:H film achieves the same decoupling but at substantially greater fabrication complexity; deployed MS filterbank spectrometers~\cite{SuperSpec_redford_2022, spt_slim, DESHIMA2} instead use deposited films, whose loss tangent has proven difficult to control. Patterning the CPW on the substrate itself reliably obtains the low loss tangent of crystalline Si without either penalty. Note that CPW structures are susceptible to radiative loss, but it can be suppressed using bridges appropriately to maintain equipotential state between the CPW ground planes.~\cite{chen2014}

In this architecture the CPW resonator is galvanically inserted between the MS feedline and its detector termination, so $Q_\mathrm{feed}$ and $Q_\mathrm{det}$ are set by the impedance mismatch at its two ends. The mismatch acts as a reflective barrier: the larger the ratio between the high-impedance CPW shunt and the low-impedance MS line, the longer power is trapped in the resonator, and the narrower the channel. This gives
\begin{align}
\label{eq:Q_impedance}
&Q_\mathrm{feed}=Q_\mathrm{det}=\frac{\pi}{2}\frac{Z_0^\mathrm{CPW}}{Z_0^\mathrm{MS}} \\
&\mathcal{R}=\left(\frac{4}{\pi}\frac{Z_0^\mathrm{MS}}{Z_0^\mathrm{CPW}}+\frac{1}{Q_\mathrm{int}}\right)^{-1},
\end{align}
derived in supplementary material Sec.~II, with analytical expressions for $Z_0^\mathrm{MS}$ and $Z_0^\mathrm{CPW}$ shown in supplementary material Sec.~III. The resolving power is therefore set by a ratio of geometric impedances, up to a ceiling imposed by $Q_\mathrm{int}$, and is tuned by adjusting the geometric parameters of the CPW and MS circuit elements.

\begin{figure}[htbp]
\includegraphics[width=\columnwidth]{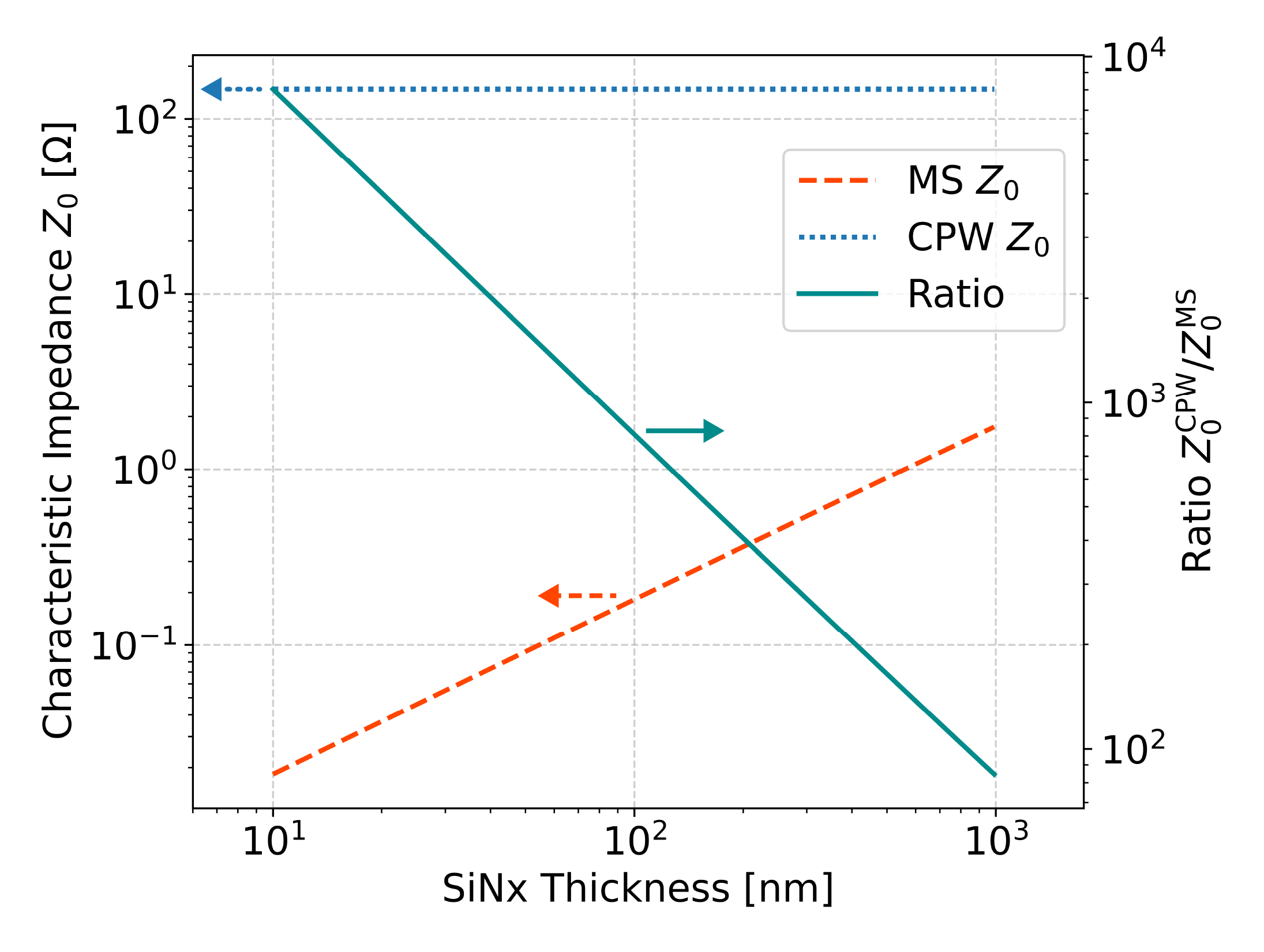}
\caption{\label{fig:geometric_tuning} Characteristic impedance $Z_0$ of MS (dashed orange) and CPW (dotted blue) and their ratio (solid cyan) as a function of Si-rich SiNx dielectric ($\epsilon_r = 8.6$) thickness. Arrows near the lines indicate their corresponding y-axes. $Z_0^\mathrm{MS}$ decreases with SiNx thickness and therefore the ratio $Z_0^\mathrm{CPW}/Z_0^\mathrm{MS}$. Because $\mathcal{R}$ scales with this ratio, decreasing the SiNx thickness leads to higher $\mathcal{R}$. A 50 nm interlayer gives an impedance ratio of $\sim1.6\times10^3$, corresponding to $\mathcal{R}\approx1000$.}
\end{figure}

Fig.~\ref{fig:geometric_tuning} illustrates the tuning mechanism. As the interlayer thins, the MS approaches the parallel-plate limit and its capacitance per unit length rises, driving $Z_0^\mathrm{MS}$ steeply down. The CPW impedance is independent of the interlayer, and is designed to reach asymptotically high impedance with a narrow signal strip and wide ground plane gap on a Si substrate. The ratio $Z_0^\mathrm{CPW}/Z_0^\mathrm{MS}$, and with it $\mathcal{R}$ through Eq.~\ref{eq:Q_impedance}, therefore rises steeply as the interlayer is thinned. For the chosen geometry, a 50 nm SiNx thickness gives $Z_0^\mathrm{MS}=0.09~\Omega$ against $Z_0^\mathrm{CPW}=148~\Omega$, yielding an impedance ratio of $1.6\times10^{3}$. With $Q_\mathrm{int}=10^{4}$ set by absorption in the Si substrate, this yields $\mathcal{R}=1140$. Reaching $\mathcal{R}\sim1000$ thus calls for an interlayer thickness of 50 nm.

\section{\label{sec:sim}Simulation Results}

Using typically achievable feature sizes for both CPW and MS geometries (detailed in supplementary material Sec.~IV), the planar structures for a ten-channel filterbank spectrometer with $\Sigma=1$, sampling between 89.6 and 90.4~GHz, were implemented in the Sonnet electromagnetic simulation environment using its Netlist feature. Note that the SiNx thickness of 50~nm is thinner than the MS interlayer dielectric commonly utilized for millimeter-wave devices. As detailed in Sec.~\ref{sec:principles}, the ten channels are differentiated solely by the physical length of their respective CPW half-wave shunts and placed at half-wave distances along the feedline.

\begin{figure}[htbp]
\includegraphics[width=\columnwidth]{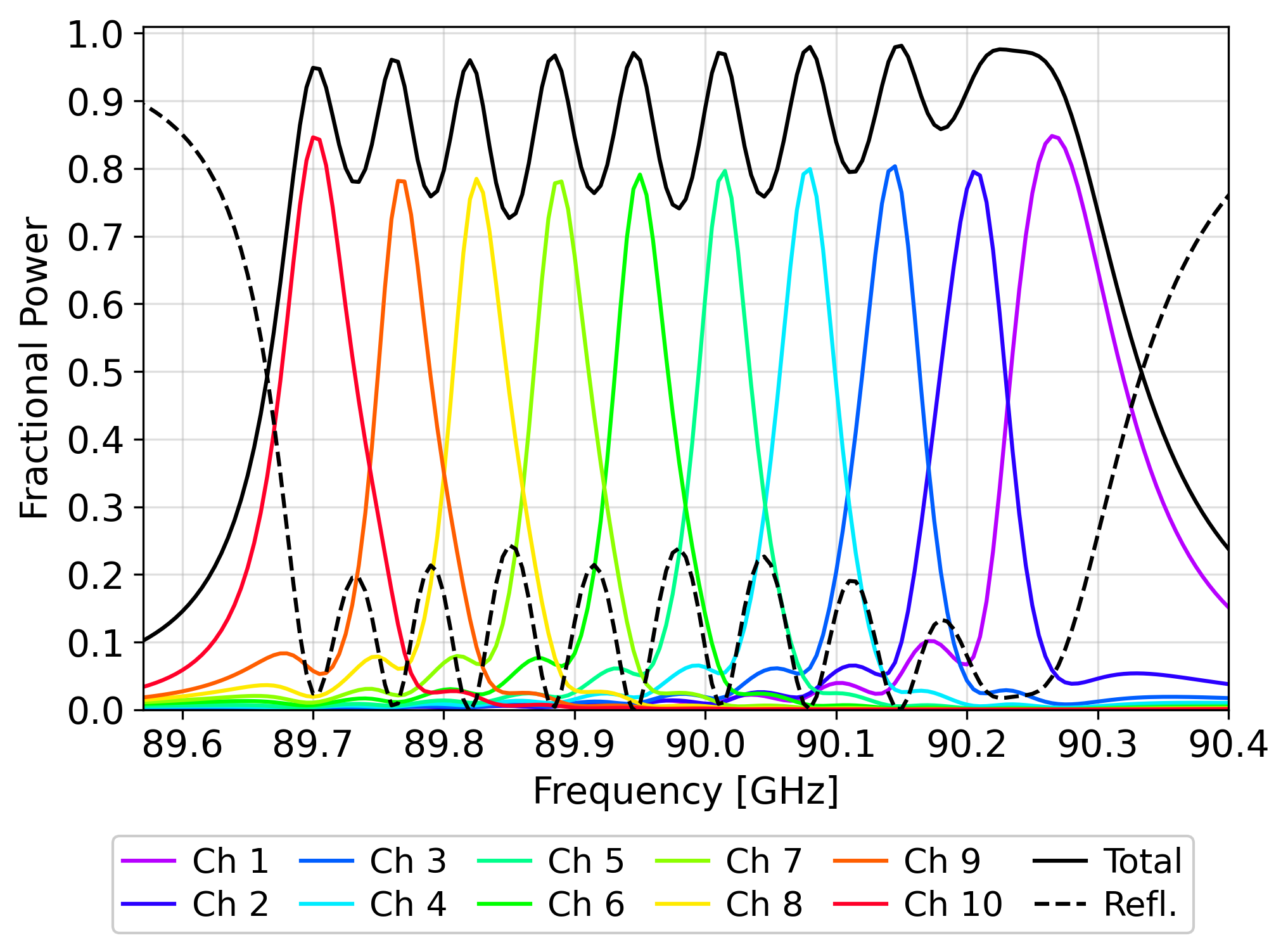}
\caption{\label{fig:Sparams} Simulated fractional power delivered for each channel (color, solid), together with the cumulative power delivered for the filterbank (black, solid) and power reflected at the feedline input (black, dashed), for a SiNx $\tan\delta=1\times10^{-3}$. Lorentzian fits to the individual channels give $\mathcal{R}=1170\pm155$. Peak channel efficiencies exceed 78\% and show no significant trend with position along the feedline.}
\end{figure}

Fig.~\ref{fig:Sparams} shows the simulated response of the filterbank with the SiNx interlayer set to $\tan\delta = 1\times10^{-3}$. Fitting a Lorentzian profile to each channel gives a resolving power of $\mathcal{R}=1170\pm155$, where quoted uncertainty is the channel-to-channel standard deviations of the fitted values, in agreement with the analytically expected value $\mathcal{R}=1140$ obtained in Sec.~\ref{sec:principles}. All peak channel efficiencies exceed 78\%, well above the 50\% ceiling of a terminated feedline, and show no significant trend from the earliest to the latest channel along the feedline architecture. Across channels 2 through 9, which are all flanked by neighbors on both sides, $\eta_\mathrm{ch}$ agrees to within 1.5\%. Channels 1 and 10 are excluded from this comparison because they are more efficient without a neighboring channel on one side. Propagation loss in the feedline is therefore negligible on the scale of this device. An independent distributed circuit of the same filterbank, implemented in Keysight Advanced Design System (ADS), reproduces this behavior (see supplementary material Sec.~V).

A comprehensive sensitivity analysis evaluating performance robustness against typical fabrication uncertainties, including CPW strip width, SiNx thickness, permittivity, and loss tangent, is detailed in supplementary material Sec.~VI.

\section{Prototype Device}

A prototype device was designed and fabricated to experimentally validate the design principles, targeting a simulated resolving power of $\mathcal{R}=92\pm5$ and channel efficiency of $\eta_\mathrm{ch}=(81\pm3)\%$ (see supplementary material Sec.~VII). Fabricated by SEEQC, Inc.~\cite{Suzuki_SEEQC}, the device chip shares wafer space with a concurrent cosmic microwave background detector fabrication run. To simplify the fabrication process and minimize changes to the existing layer stack, the prototype adopts the wafer's existing 560~nm SiNx interlayer, in place of the 50~nm layer modeled in Sec.~\ref{sec:sim}, and the CPW traces are patterned on top of SiNx rather than the Si wafer. The thicker SiNx raises $Z_0^\mathrm{MS}$ and lowers the impedance ratio $Z_0^\mathrm{CPW}/Z_0^\mathrm{MS}$, scaling $\mathcal{R}$ down by an order of magnitude. Patterning the CPW traces on the SiNx layer rather than directly on single-crystal Si minimizes changes to the layer stack but increases the effect of absorption loss per channel by $\sim7\%$. Consequently, while evaluating $Q_\mathrm{int}$ decoupling from interlayer loss requires the Nb-on-Si geometry of Sec.~\ref{sec:sim}, this prototype simultaneously validates two core principles of the device architecture: that removing the feedline termination achieves $\eta_\mathrm{ch} > 50\%$, and that $\mathcal{R}$ scales with the impedance ratio.

\begin{figure}[htbp]
    \centering
    \includegraphics[width=0.95\columnwidth]{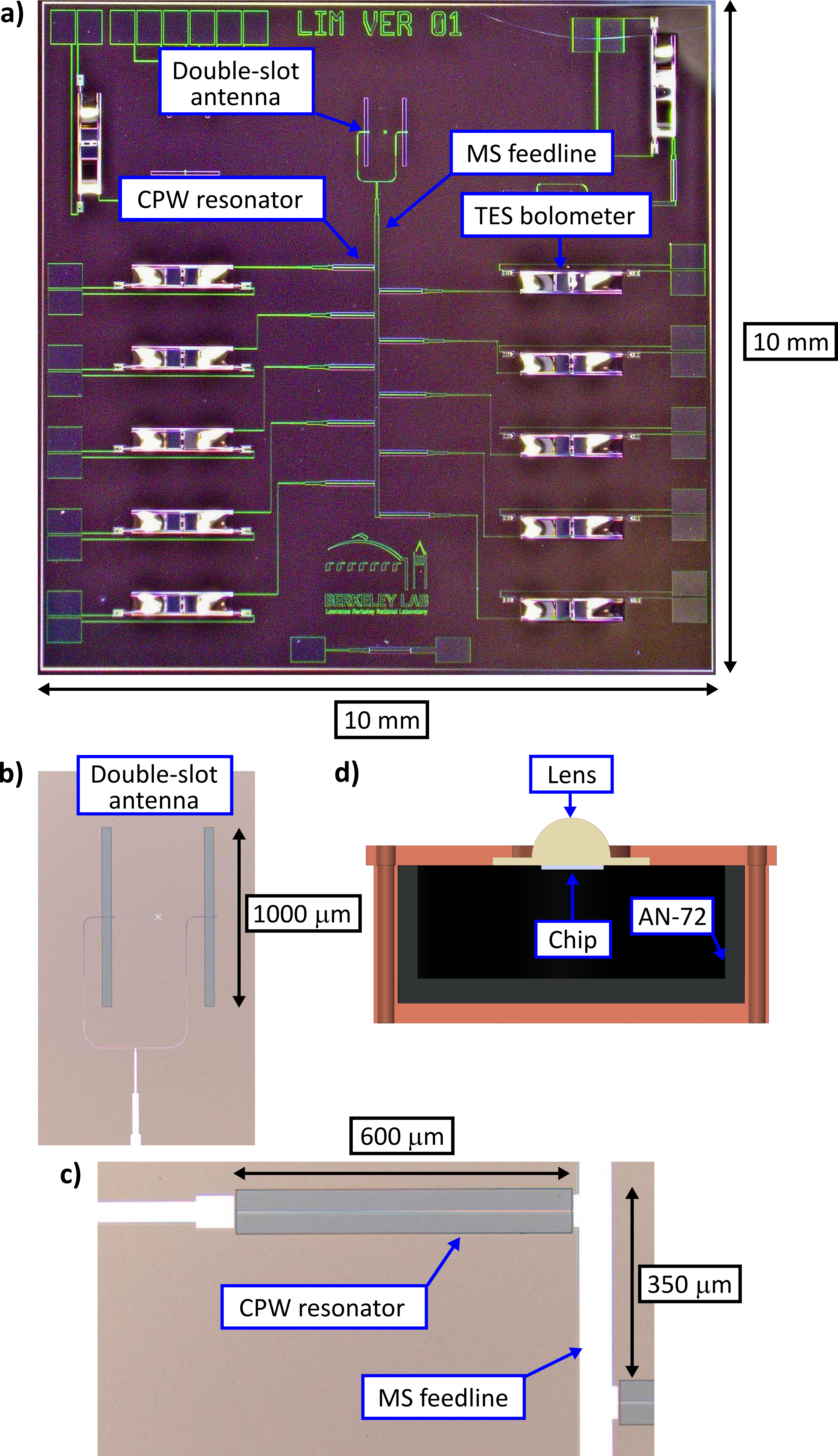}
    \caption{(a) Photograph of the prototype device, with closeups of the (b) double-slot antenna and (c) MS feedline and CPW resonator shunt. The MS feedline consists of a Nb signal line and ground plane separated by a 560~nm SiNx dielectric interlayer. To simplify fabrication, the CPW resonator's Nb signal trace and ground planes are patterned directly on top of the SiNx dielectric layer. Shown in the cross-sectional schematic (d), the chip is mounted on its backside to a 12.7~mm diameter alumina hemispherical lens while its front side faces a cold Eccosorb AN-72 cavity. This entire setup is mounted to the cold stage of a DR and cooled to 250 mK.}
    \label{fig:chip_schematic}
\end{figure}

The fabricated device chip is shown in Figs.~\ref{fig:chip_schematic}(a)--(c). It features a double-slot antenna for free-space coupling, an MS feedline with Nb signal trace and ground plane separated by a 560~nm SiNx dielectric interlayer, CPW resonator shunts, and a titanium (Ti) transition-edge sensor (TES) bolometer for each filterbank channel. The TES is described in more detail by Rotermund \textit{et al}.~\cite{rotermund2024} As shown in Fig.~\ref{fig:chip_schematic}(d), the antenna is centered on an extended hemispherical alumina lens of 12.7 mm diameter which is anti-reflection coated with thermoformed polyetherimide. This lens focuses light onto the antenna from the back-side of the device while the front-side of the device faces an Eccosorb AN-72 cavity acting as a backlobe absorber. The TES is voltage-biased into the superconducting transition and its current-changes readout using a superconducting quantum interference device (SQUID).

The device is anchored to the cold stage of a dilution refrigerator (DR) and operated at 250 mK, where the Ti TES is voltage-biased into its superconducting transition. To illuminate the device with external signal for characterization, a laminated Zotefoam HD30 window is integrated into the vacuum shell. To reduce out-of-band infrared loading on the cold stages, a series of optical low-pass filters is installed between the vacuum window and the 1 K stage of the DR. From warm to cold, the stack comprises two 3.2~mm expanded PTFE layers and a 12.7~mm solid PTFE filter at 50~K, $12\text{ cm}^{-1}$ and $8.5\text{ cm}^{-1}$ metal-mesh filters at 4~K, and a final $6.5\text{ cm}^{-1}$ metal-mesh filter at 1~K. The stack was characterized independently using the heterodyne setup described by Jeong $\textit{et al}$.~\cite{jeong_AR_2023}

\begin{figure*}[htbp]
    \centering
    \includegraphics[width=0.85\linewidth]{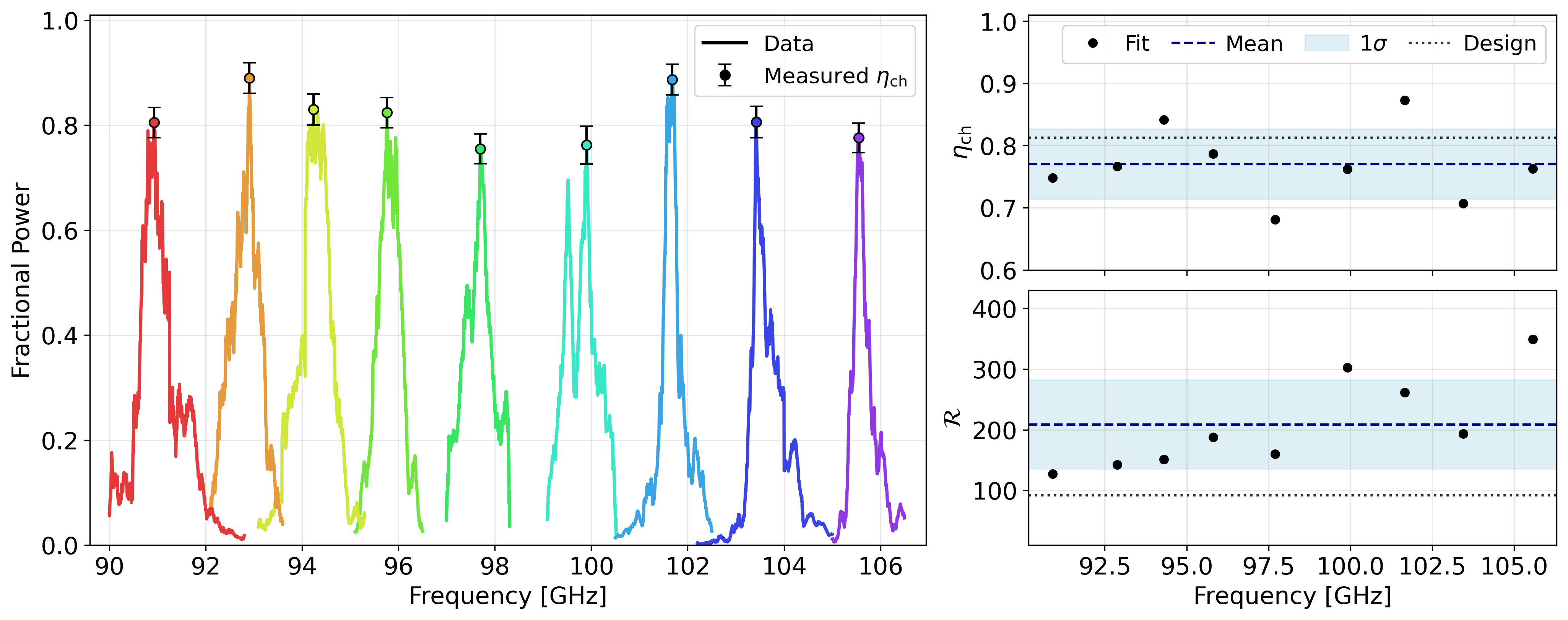}
    \caption{Measurement results showing the efficiency normalized spectral profile of the prototype chip channels. Fits to these curves yield $\mathcal{R}=208\pm74$ and $\eta_\mathrm{ch}=(77\pm6)\%$, where the uncertainties are the channel-to-channel standard deviations. Black dotted lines mark the simulated design values of $\mathcal{R}=92$ and $\eta_\mathrm{ch}=81\%$.}
    \label{fig:measurement}
\end{figure*}

To measure the filterbank frequency response, a frequency-tunable continuous wave (CW) source illuminates the hemispherical lens through the vacuum window. A lock-in amplifier modulates the source signal and synchronously demodulates the resulting SQUID readout. To determine $\eta_\mathrm{ch}$, the device views a 77 K liquid nitrogen source through a 12.7 mm diameter aperture in an Eccosorb AN-72 absorber sheet. A room-temperature chopper wheel coated with Eccosorb AN-72 periodically alternates the incident radiation between the 77 K and the 300 K loads. Detailed schematics of both test configurations are provided in Supplementary Material Sec.~VIII. For a TES operating under strong negative electrothermal feedback, the change in electrical bias power tracks the change in coupled optical power,
\begin{equation}
\label{eq:eta_extract}
\Delta P_\mathrm{elec} = -\eta_\mathrm{opt} \eta_\mathrm{ch}  \gamma k_B  \Delta T   \Delta\nu_\mathrm{eff},
\end{equation}
where $\Delta T$ is the physical temperature difference between the two loads, and $\Delta\nu_\mathrm{eff} = \int f(\nu)\,d\nu$ is the effective channel bandwidth integrated over the normalized spectral response $f(\nu)$ measured with the CW source. To separate spatial beam geometry from optical path loss, $\gamma = \int_\mathrm{aperture} G(\theta,\phi)\,d\Omega / \int_{0}^{4\pi} G(\theta,\phi)\,d\Omega$ represents the dimensionless geometric beam-coupling factor (the fraction of the lens-coupled antenna gain pattern $G(\theta,\phi)$ subtended by the calibration aperture), whereas $\eta_\mathrm{opt}$ represents the net optical transmission efficiency of everything preceding the filterbank. This term accounts for loss through the vacuum window, filter stack, and anti-reflection coated lenslet, as well as the electromagnetic coupling between the lenslet and antenna, the impedance mismatch between the antenna slot and MS, and propagation along the MS line through the impedance transformer to the filterbank input feedline. The remaining term, $\eta_\mathrm{ch}$, is the channel efficiency from the filterbank feedline input to the TES detector. Evaluating Eq.~\ref{eq:eta_extract} yields the combined product $\eta_\mathrm{opt}\eta_\mathrm{ch}$, from which $\eta_\mathrm{ch}$ is isolated by dividing out $\eta_\mathrm{opt}$, which is determined by combining measured cryostat filter transmissions with electromagnetic simulations of the lenslet-coupled antenna and feedline performed in Ansys HFSS and Sonnet. 

Fig.~\ref{fig:measurement} shows the measured performance, combining the spectral response profile from the CW source with the $\eta_\mathrm{ch}$ calibration from the chopped liquid nitrogen source. Note that the highest frequency channel (Ch 1) was unresponsive due to an open circuit in its shunt, hence the nine remaining channels are characterized here. The total uncertainty in $\eta_\mathrm{ch}$ is computed by adding the errors in $B_\nu^\mathrm{eff}$, $\Delta P_\mathrm{elec}$, and $\gamma$ in quadrature. Errors in $B_\nu^\mathrm{eff}$ and $\Delta P_\mathrm{elec}$ represent the standard error of the mean over repeated trials, while the error in $\gamma$ is propagated from a $\pm5$~mm source position uncertainty. Fitting a Lorentzian profile to the measured curves yields $\mathcal{R} = 208 \pm 74$ and $\eta_\mathrm{ch} = (77 \pm 6)\%$, where the uncertainties are the channel-to-channel standard deviations. The measured efficiency agrees with the simulated $(81\pm3)\%$ to within the uncertainty and lies well above the 50\% ceiling of a terminated feedline. The measured resolving power exceeds the prediction by roughly a factor of two, and the measured center frequencies sit $(4.5\pm 0.2)\%$ above the design. Both offsets are under investigation, as is a possible trend of $\mathcal{R}$ with position along the feedline. Notwithstanding these offsets, the measurement confirms the two design principles: efficiency well above the terminated limit, and a resolving power set by the impedance ratio between the feedline and the shunts.

\section{Conclusion}
We have presented a superconducting filterbank architecture defined by two design choices: eliminating the feedline termination resistor and implementing CPW resonators on a low-loss single-crystal Si substrate. In this design, the resolving power is set by the ratio of the resonator and feedline characteristic impedances. Electromagnetic and circuit simulations of a ten-channel filterbank near 90~GHz confirm its high efficiency, medium resolution design, giving $\mathcal{R}=1170$ with $\eta_\mathrm{ch}> 78\%$ and no significant trend in efficiency across the channels. A prototype device, designed for $\mathcal{R}=92\pm5$ and $\eta_\mathrm{ch}=(81\pm3)\%$ based on simulations, measured $\mathcal{R}=208\pm74$ and $\eta_\mathrm{ch}=(77\pm6)\%$, validating both the removal of the 50\% efficiency ceiling and the tuning of $\mathcal{R}$ through the circuit impedance ratio. While an MS filterbank on a single-crystal or a low-loss dielectric film interlayer would achieve comparable $\mathcal{R}$ and $\eta_\mathrm{ch}$, doing so requires extensive process development, such as a SOI fabrication flow for single-crystal Si interlayer. in contrast, patterning the resonator on the Si substrate obtains the same low loss from the wafer itself. In our prototype, the measured $\mathcal{R}$ exceeds simulation by roughly a factor of two and the center frequencies sit $(4.5\pm 0.2)\%$ above design. Future work will resolve the origin of these offsets and will further demonstrate the scalability of the architecture for $\mathcal{R}>1000$ and high channel count required for next-generation millimeter-wave spectroscopic surveys.

\section*{Supplementary Material}
See the Supplementary Material for additional circuit models, impedance derivations, simulation stackup parameters, fabrication tolerance studies, and detailed measurement setup descriptions.

\begin{acknowledgments}
The authors thank Adrian T. Lee, Martin White, and Kaja M. Rotermund for useful discussions. This work is supported by the U.S. Department of Energy, Office of Science, Office of High Energy Physics Instrumentation and Detector R\&D under Contract Numbers DE-AC02-05CH11231.
\end{acknowledgments}

%\nocite{*}
\bibliography{aipsamp}

% ==============================================================================
% START OF SUPPLEMENTARY MATERIAL
% ==============================================================================
% Reset counters to zero so supplementary figures and equations start at 1
\setcounter{section}{0}
\setcounter{figure}{0}
\setcounter{table}{0}
\setcounter{equation}{0}
% Redefine prefixes to add 'S'
\renewcommand{\thefigure}{S\arabic{figure}}
\renewcommand{\thetable}{S\arabic{table}}
\renewcommand{\theequation}{S\arabic{equation}}
\renewcommand{\thesection}{\Roman{section}}

\section*{Supplementary Material}

\section{Circuit Model Comparison of Terminated vs. Unterminated Feedlines}

Conventional on-chip filterbanks employ a resistive terminator after the final channel to mitigate standing waves and residual reflections.~\cite{Barry} However, this design choice immediately imposes a fundamental efficiency cap of $\eta_\mathrm{ch} = 50\%$ per channel, even in the ideal, lossless limit.~\cite{hd_superspec}

\begin{figure*}[htbp]
\centering
\includegraphics[width=0.73\textwidth]{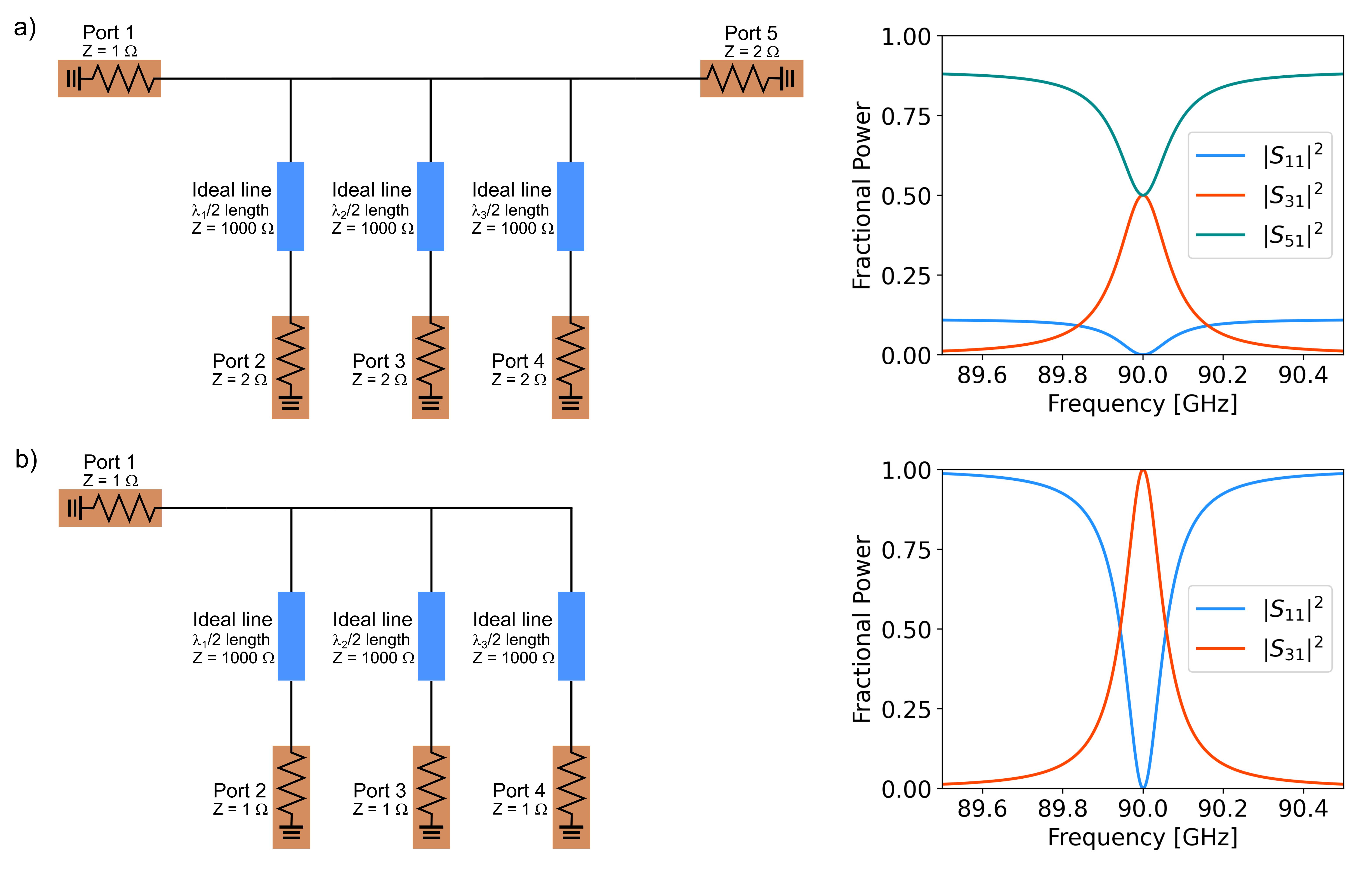}
\caption{\label{fig:ADS}Circuit model and resulting filter efficiency using ADS for (a) a representative circuit of existing technologies with a termination resistor and (b) the circuit principle employed for this technology without a termination resistor. While (a) shows that power is split evenly between the filter and termination resistor at the resonant frequency with $|S_{31}|^2$ capped at 0.5, (b) shows that all of the power is transmitted through the filter with $|S_{31}|^2$ = 1. Port impedances are set for maximal power transmitted to Port 3.}
\end{figure*}

Under ideal conditions, Fig.~\ref{fig:ADS}a demonstrates that conventional terminated filterbanks split power evenly between the resonant filter and the load resistor at resonance, bounding the peak transmission to $|S_{21}|^2 = 0.5$. In contrast, Fig.~\ref{fig:ADS}b shows that eliminating the termination resistor removes this loss mechanism, allowing peak power transmission to reach $|S_{21}|^2 = 1.0$.

\section{Derivation of Resolving Power vs.\ Impedance Ratio}
\label{supp:derivation}

Each channel consists of a half-wave CPW section of characteristic impedance $Z_0^\mathrm{CPW}$, physical length $l$, and complex propagation constant $\gamma = \alpha + j\beta$, where $\alpha$ is the attenuation constant and $\beta$ is the phase constant. The CPW section is tapped galvanically to the MS feedline at one end and terminated at the detector at the other. Because $Z_0^\mathrm{CPW} \gg Z_0^\mathrm{MS}$, both ends approximate short circuits as seen from inside the resonator, establishing a series resonance at the angular frequency $\omega_0$ where $\beta l = \pi$.

The impedance looking into a short-circuited line of length $l$ is $Z_\mathrm{in} = Z_0^\mathrm{CPW} \tanh(\gamma l)$. Expressing the frequency near resonance as $\omega = \omega_0 + \delta\omega$ with $\beta l = \pi(1 + \delta\omega/\omega_0)$ for a non-dispersive line, and expanding for low loss ($\alpha l \ll 1$) and small detuning ($\delta\omega/\omega_0 \ll 1$):
\begin{equation}
\label{eq:supp_expand}
Z_\mathrm{in} = Z_0^\mathrm{CPW} \frac{\tanh\alpha l + j\tan\beta l}{1 + j\tanh\alpha l \tan\beta l} \;\simeq\; Z_0^\mathrm{CPW}\left(\alpha l + j\pi\frac{\delta\omega}{\omega_0}\right).
\end{equation}
This matches the input impedance of a series $RLC$ circuit, $Z \simeq R_\mathrm{loss} + 2jL\delta\omega$, with equivalent circuit elements given by
\begin{equation}
\label{eq:supp_rlc}
R_\mathrm{loss} = Z_0^\mathrm{CPW}\alpha l, \qquad L = \frac{\pi Z_0^\mathrm{CPW}}{2\omega_0}, \qquad C = \frac{1}{\omega_0^2 L}.
\end{equation}
The equivalent series inductance $L$ is fixed solely by the resonator's characteristic impedance, establishing the origin of the impedance ratio scaling.

The internal quality factor $Q_\mathrm{int}$ is determined by the intrinsic line loss:
\begin{equation}
\label{eq:supp_qint}
Q_\mathrm{int} = \frac{\omega_0 L}{R_\mathrm{loss}} = \frac{\pi}{2\alpha l} = \frac{\beta}{2\alpha},
\end{equation}
which collects conductor, dielectric, and radiative dissipation. For a CPW on a single-crystal Si substrate, $\alpha$ is dominated by the substrate and residual slotline mode radiation rather than a deposited dielectric interlayer, decoupling $Q_\mathrm{int}$ from SiNx material loss.

The external terminations at the feedline and detector appear in series with this equivalent circuit. Denoting $Z_\mathrm{ext}$ as the real part of the impedance presented to either end, each port contributes an external quality factor $Q_\mathrm{ext} = \omega_0 L / Z_\mathrm{ext} = (\pi/2) (Z_0^\mathrm{CPW} / Z_\mathrm{ext})$. Assuming a traveling-wave feedline ($Z_\mathrm{ext} = Z_0^\mathrm{MS}$) matched to the detector termination:
\begin{equation}
\label{eq:supp_qfeed}
Q_\mathrm{feed} = Q_\mathrm{det} = \frac{\pi}{2} \frac{Z_0^\mathrm{CPW}}{Z_0^\mathrm{MS}}.
\end{equation}
The overall channel resolving power $\mathcal{R}$ is then obtained by combining internal and external loss rates:
\begin{equation}
\label{eq:supp_R}
\mathcal{R} = Q = \left(\frac{1}{Q_\mathrm{feed}} + \frac{1}{Q_\mathrm{det}} + \frac{1}{Q_\mathrm{int}}\right)^{-1} = \left(\frac{4}{\pi}\frac{Z_0^\mathrm{MS}}{Z_0^\mathrm{CPW}} + \frac{1}{Q_\mathrm{int}}\right)^{-1}.
\end{equation}
Equation~\ref{eq:supp_R} demonstrates that $\mathcal{R}$ scales directly with the geometric impedance ratio $Z_0^\mathrm{CPW} / Z_0^\mathrm{MS}$ until coupling loss approaches intrinsic loss, beyond which performance transitions from geometry-limited to loss-limited.

\section{Transmission Line Characteristic Impedance Expressions}

The characteristic impedance of the MS feedline is given by:~\cite{pozar}
\begin{equation}
    \label{eq:ms_z0}
    Z_0^\mathrm{MS}=\frac{120\pi}{\sqrt{\epsilon_\mathrm{eff}^\mathrm{MS}}\left(w/t+1.393+0.667\log\left(w/t+1.444\right)\right)},
\end{equation}
where $w$ and $t$ are the trace width and dielectric thickness, and the effective relative permittivity is
\begin{equation}
    \label{eq:ms_epsilon}
    \epsilon_\mathrm{eff}^\mathrm{MS}=\frac{\epsilon_r+1}{2}+\frac{\epsilon_r-1}{2\sqrt{1+12t/w}}.
\end{equation}

Similarly, the characteristic impedance for CPW is given by:~\cite{simons}
\begin{equation}
    \label{eq:CPW_z0}
    Z_0^\mathrm{CPW}=\frac{30\pi}{\sqrt{\epsilon_\mathrm{eff}^\mathrm{CPW}}}\frac{K(k_0^\prime)}{K(k_0)},
\end{equation}
where $K(k)$ is the complete elliptical integral of first kind and $\epsilon_\mathrm{eff}^\mathrm{CPW}$ is the effective relative permittivity described by
\begin{equation}
    \label{eq:CPW_epsilon}
    \epsilon_\mathrm{eff}^\mathrm{CPW}=\frac{1+\epsilon_r}{2},
\end{equation}
with $k_0 = \frac{s}{s+2d}$ and $k_0^\prime=\sqrt{1-k_0^2}$. Here, $s$ is the width of the signal line trace and $d$ is the width of the gap between this trace and the nearest ground plane edge.

\section{Sonnet Electromagnetic Simulation Parameters}

\begin{figure}[htbp]
\centering
\includegraphics[width=0.5\linewidth]{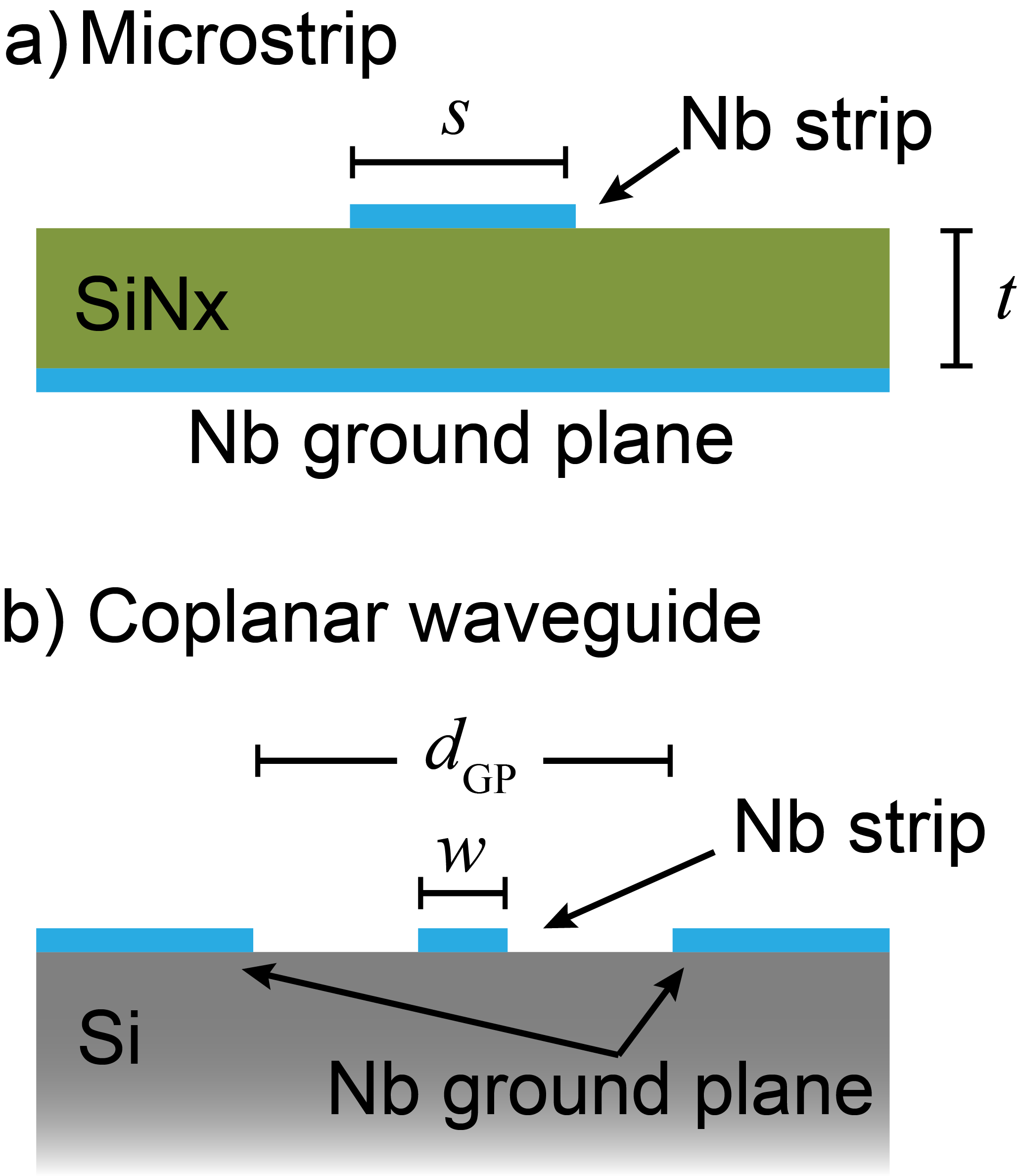}
\caption{\label{fig:MS_CPW} Cross-sectional illustration of the (a) MS and (b) CPW architectures utilized in this filterbank design. Labels correspond to the design parameters and material layers detailed in Tab.~\ref{tab:sim_sizes}.}
\end{figure}

Tab.~\ref{tab:sim_sizes} lists the design parameters and nominal values implemented in the Sonnet electromagnetic simulation environment. Fig.~\ref{fig:MS_CPW} provides a cross-sectional illustration of the MS feedline and CPW resonator geometries.

\begin{table}[htbp]
\caption{\label{tab:sim_sizes}Design parameters and nominal values for the filterbank spectrometer as implemented in Sonnet EM simulations. Refer to Fig.~\ref{fig:MS_CPW} for a cross-sectional illustration showing these parameters.}
\begin{ruledtabular}
\begin{tabular}{llc}
\textbf{Component} & \textbf{Parameter} & \textbf{Value} \\
\hline
\textbf{CPW Resonator} & Signal strip width ($s$) & 2~$\mathrm{\mu}$m \\
                       & Ground plane gap width ($d_\mathrm{GP}$)           & 80~$\mu$m \\
                       & Si relative permittivity & 11.7 \\
                       & Si loss tangent~\cite{Si_loss} & $1 \times 10^{-5}$ \\
\hline
\textbf{MS Feedline}   & Strip width ($w$)        & 70~$\mu$m \\
                       & SiNx thickness ($t$) & 50~nm \\
                       & SiNx relative permittivity\footnotemark[1] & 8.6 \\
                       & SiNx loss tangent & $1 \times 10^{-3}$ \\
\hline
\textbf{Nb Superconductor} & Surface inductance ($L_s$)\footnotemark[2] & 0.15~pH/$\square$\footnotemark[3] \\
\end{tabular}
\end{ruledtabular}
\footnotetext[1]{High relative permittivity SiNx is developed by Suzuki \textit{et al} for cosmic microwave background measurement applications.}
\footnotetext[2]{Surface inductance is utilized within the Sonnet environment to account for the kinetic inductance of the superconducting thin film.}
\footnotetext[3]{Measured using resonator at 6 GHz by the Lawrence Berkeley National Laboratory Superconducting Detector Group.}
\end{table}

\section{Distributed circuit model of the $\mathcal{R}\approx1100$ filterbank}
\begin{figure*}[htbp]
\centering
\includegraphics[width=0.68\linewidth]{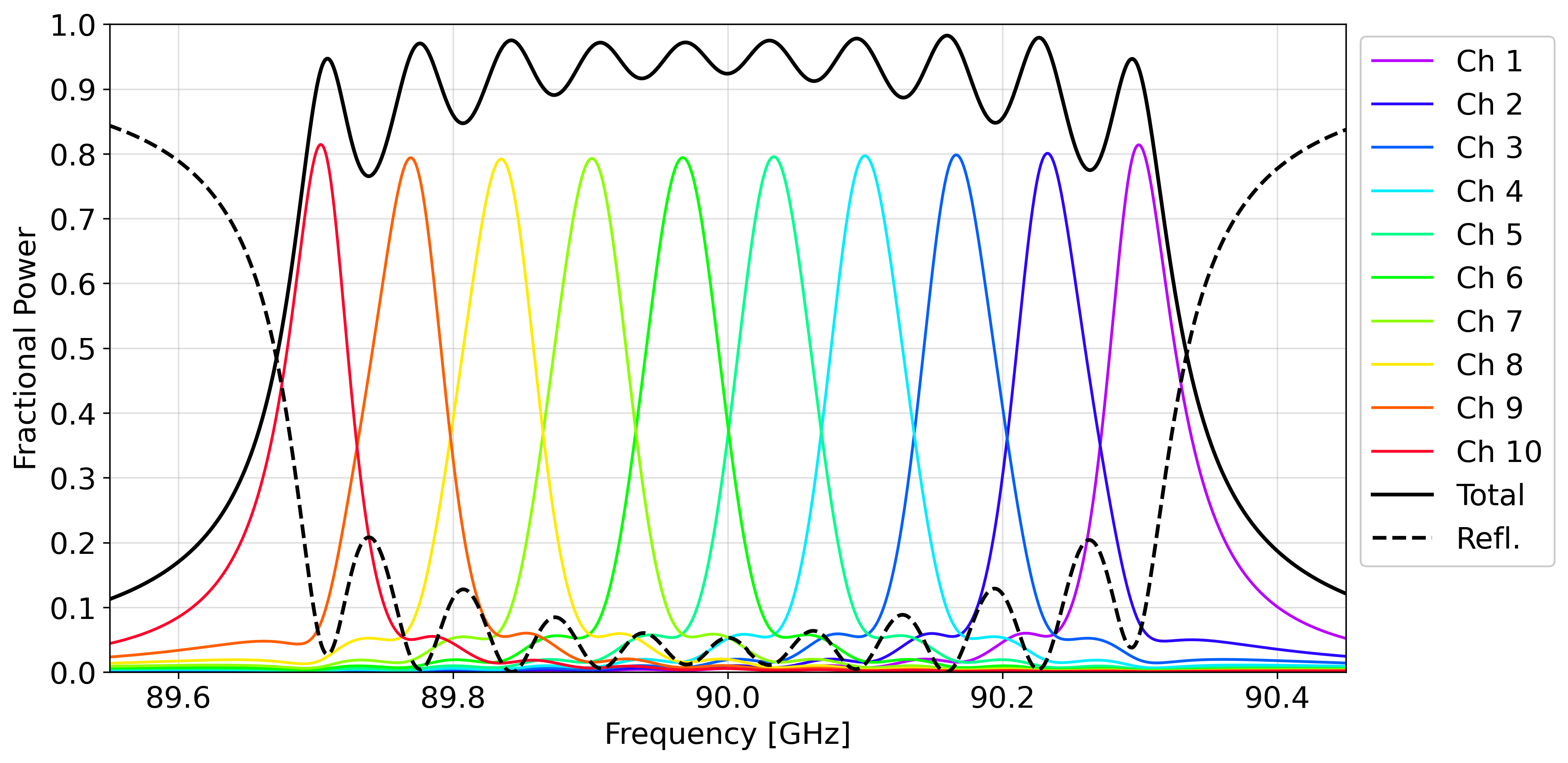}
\caption{\label{fig:ADS_10ch} Distributed circuit model of the ten-channel bank in ADS, with exact channel frequencies and shunt spacings at unity oversampling and $\tan\delta=1\times10^{-3}$. Shown are the simulated fractional power delivered to each channel (color, solid), cumulative power delivered for the filterbank (black, solid), and power reflected at the feedline input (black, dashed). Lorentzian fits give $\mathcal{R}=1095\pm120$ and $\eta_\mathrm{ch}=(80.1\pm0.7)\%.$}
\end{figure*}

The electromagnetic simulation of Sec.~III of the main text is cross-checked against a distributed circuit model of the same ten-channel filterbank, implemented in ADS. Each channel is represented by a transmission line of the designed length and characteristic impedance, tapped onto the transmission line and terminated on a matched resistive load standing in for the detector; the feedline is modeled as a lossy transmission line carrying the loss tangent $\tan\delta=1\times10^{-3}$ of the SiNx interlayer. Unlike the Sonnet simulation environment, the circuit model does not place any constraint on the discretization of lengths, so channel frequencies and shunt spacings are set at their exact values at unity oversampling.
Fig.~\ref{fig:ADS_10ch} shows the resulting fractional power delivered to each detector port. Lorentzian fits give $\mathcal{R}=1095\pm120$ and $\eta_\mathrm{ch}=(80.1\pm0.7)\%$, where the uncertainties are the 1-$\sigma$ spread across the ten channels, in agreement with the Sonnet simulation. The peak efficiencies between the second and ninth channels differ by less than 0.7\%, confirming that loss along the feedline is negligible on the scale of this device.

\section{Fabrication Tolerance and Sensitivity Analysis}

\begin{figure*}[htbp]
\centering
\includegraphics[width=0.79\textwidth]{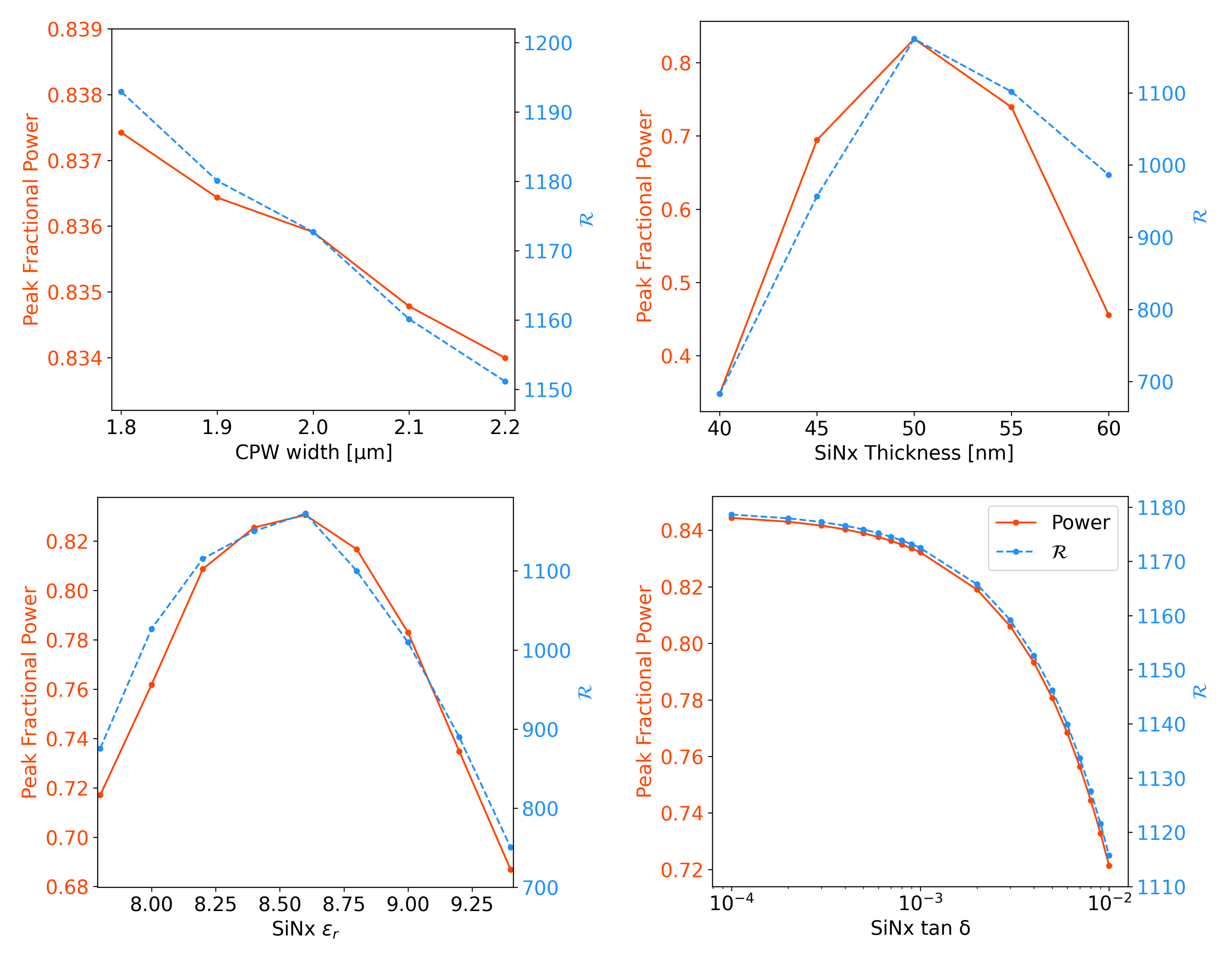}
\caption{\label{fig:tolerance}Sensitivity study of the middle channel of a representative 5-channel sub-unit. The dependence of $\eta_\mathrm{ch}$ (solid orange) and $\mathcal{R}$ (dashed blue) on design parameters with the largest fractional uncertainties is studied by performing a parameter sweep in Sonnet. The design parameters studied are the width of the CPW signal line (top-left), SiNx thickness (top-right), SiNx relative permittivity (bottom-left), and SiNx loss tangent (bottom-right). Arrows near the lines indicate their corresponding y-axes.}
\end{figure*}

The robustness of the filterbanks against fabrication uncertainties is evaluated by performing a parameter sweep using a representative 5-channel sub-unit. To characterize potential performance degradation, the dependence of $\eta_\mathrm{ch}$ and $\mathcal{R}$ for the middle channel on the design parameters with the largest fractional uncertainties is examined - the width of the CPW signal strip $s$, the MS SiNx thickness $t$, SiNx $\epsilon_r$, and SiNx $\tan\delta$. While all parameters are varied by $\pm10\%$, SiNx $\tan\delta$ is varied between $10^{-4}$ to $10^{-2}$ to account for possible degradation seen by Benson \textit{et al.}~\cite{spt_slim} Fig.~\ref{fig:tolerance} shows the peak $\eta_\mathrm{ch}$ and $\mathcal{R}$ for the channel with respect to variations from the fiducial values of these design parameters.

The analysis demonstrates that the spectrometer performance is largely insensitive to variations in CPW signal strip $s$, SiNx $\epsilon_r$, and SiNx $\tan\delta$. However, deviations exceeding 5\% in SiNx thickness causes a sharp decline in both $\eta_\mathrm{ch}$ and $\mathcal{R}$. This sensitivity arises from the strong dependence of the MS electromagnetic properties, and consequently the $\lambda/2$ shunt spacing scheme, on the dielectric thickness. A similar mechanism drives the system's sensitivity to the SiNx $\epsilon_r$, which represents the next most significant design dependency. Because shifts in $\epsilon_r$ also disrupts the phase-matching of the shunt intervals, a 10\% increase in relative permittivity results in a 17\% relative reduction in peak $\eta_\mathrm{ch}$ and a 36\% decrease in $\mathcal{R}$. However, even under such extreme degradation, both $\eta_\mathrm{ch}$ and $\mathcal{R}$ remain high, indicating that the architecture possesses sufficient inherent performance margin to absorb substantial deviation in SiNx $\epsilon_r$ from the nominal value. 

Alternatively, rather than compromising the design targets, atomic layer deposition (ALD) offers a compelling alternative to plasma-enhanced chemical vapor deposition (PECVD) typically used for fabricating thin dielectric interlayers of CMB detectors. While PECVD-based methods often struggle with precision and reliability at thicknesses near 50 nm, ALD provides the atomic-scale control required to mitigate performance degradation. By leveraging a self-limiting growth mechanism with deposition rates routinely below 1 {\AA} per cycle, it enables sub-percent thickness variation for a 50 nm dielectric while simultaneously ensuring the fabrication of inherently pinhole-free films. Furthermore, ALD is capable of depositing high permittivity, low loss dielectrics such as aluminum oxide (Al$_2$O$_3$). However, if ALD is adopted, the dielectric $\tan\delta$ of ALD-produced films must be rigorously characterized to ensure target efficiencies are met.

Notably, the channel $\mathcal{R}$ remains insensitive to degradation in the SiNx $\tan\delta$; even a tenfold increase in $\tan\delta$ to $1\times10^{-2}$ results in only a 5\% reduction in $\mathcal{R}$, providing a significant safety margin against possible runaway increase in SiNx $\tan\delta$ that has historically hindered existing technologies. However, a 13\% decrease in peak $\eta_\mathrm{ch}$ for a single channel fails to capture the cumulative impact on the feedline channel count, where small per-channel losses compound rapidly in high channel-count filterbank arrays. 

Ultimately, this architecture proves highly robust against most critical fabrication uncertainties because the primary resonant structure - the CPW with Nb metalization directly on low loss Si substrate - is decoupled from the lossier dielectric layer. While the design is robust, the integrity of the CPW interface must be considered; given the sensitivity of resonator $Q$ on trench depth within Si of the CPW, there is risk of performance degradation due to over-etching into Si during Nb patterning. To ensure robustness of the Nb-on-Si resonator performance, future work will explore high selectivity etching processes such as optimized plasma chemistry or wet etch techniques, and the adoption of low loss single-crystal sapphire substrates which offer excellent chemical resistance against over-etching.

\section{Simulated performance of the prototype device}

\begin{figure*}[htbp]
\centering
\includegraphics[width=0.68\linewidth]{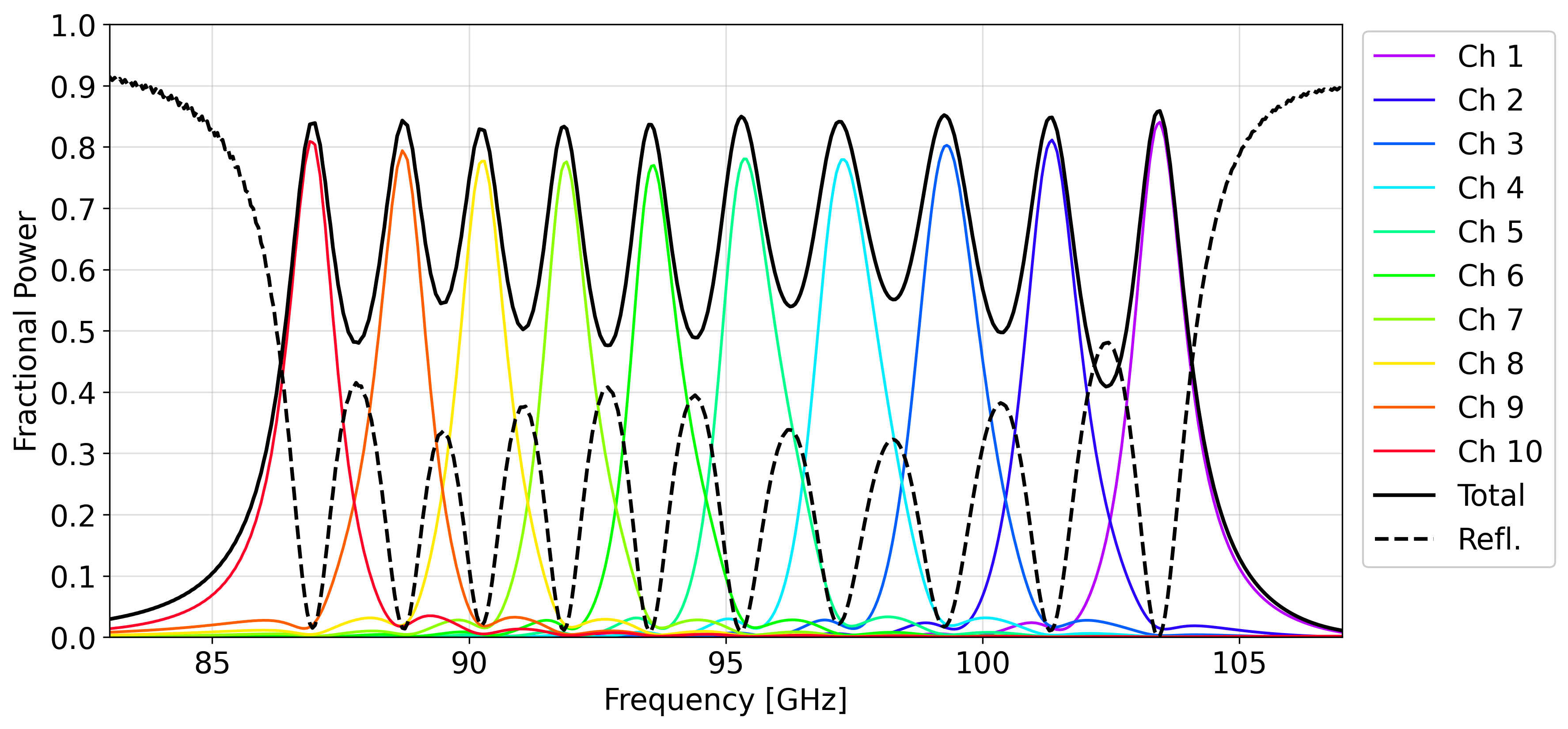}
\caption{\label{fig:prototype_sim} Simulated channel response of the prototype device in Sonnet with 560~nm Si-rich $\text{SiN}_x$ interlayer thickness, showing fractional power delivered to each channel (color, solid), cumulative power delivered for the filterbank (black, solid), and power reflected at the feedline input (black, dashed). Lorentzian fits give $\mathcal{R}=92\pm5$ and $\eta_\mathrm{ch}=(81\pm3)\%.$}
\end{figure*}

The prototype fabricated by SEEQC, Inc.\ differs from the design of Sec.~III of the main text in two respects, each of which lowers the resolving power. Because the chip shared wafer space with a concurrent cosmic microwave background detector fabrication run, the prototype adopted the wafer's existing 560~nm Si-rich SiNx interlayer to simplify fabrication and minimize changes to the established layer stack. This raises $Z_0^\mathrm{MS}$ and correspondingly reduces the impedance ratio that sets $Q_\mathrm{feed}$ and $Q_\mathrm{det}$. The CPW signal trace and ground planes are patterned on top of the SiNx interlayer rather than directly on single-crystal Si, a simplification adopted to reduce the process complexity of the first fabrication run; this increases per-channel absorption loss by $\sim 7\%$ and the internal quality factor falls from $Q_\mathrm{int}\approx10^4$ to $\approx10^3$, which by Eq.~\ref{eq:supp_R} contributes directly to $1/\mathcal{R}$. The channel dimensions were additionally chosen to place the band at 90--106~GHz with $\Sigma\approx0.6$, undersampling the band for ease of characterization. Together these give a design target of $\mathcal{R}\approx100$. 

Fig.~\ref{fig:prototype_sim} shows the simulated response of the ten channels in Sonnet, with the geometry and material parameters of Tab.~\ref{tab:sim_sizes} with the exception of SiNx thickness $t=560$ nm and the Nb trace of the CPWs on top of SiNx. Lorentzian fits give $\mathcal{R}=92\pm5$ and $\eta_\mathrm{ch}=(81\pm3)\%$, where the quoted uncertainties are the channel-to-channel standard deviations of the fitted values. Note that while individual channel efficiencies are high, overall cumulative efficiency is reduced by lower channel overlap and the $\sim 7\%$ absorption loss penalty. This simulation provides the direct prediction against which the measurements of Sec.~IV of the main text are compared.

\section{Measurement setup}

\begin{figure}[h!]
\centering
\includegraphics[width=0.9\columnwidth]{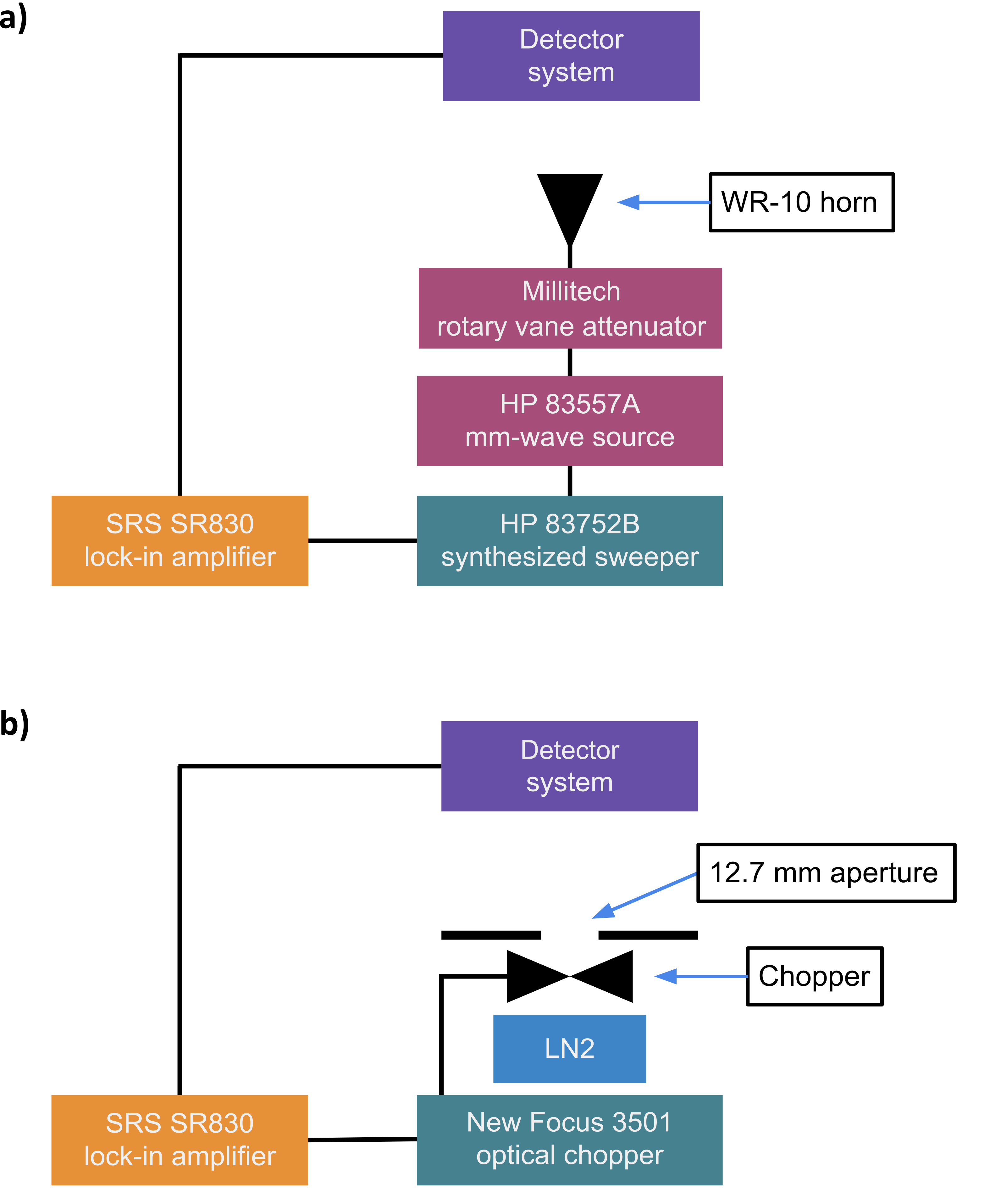}
\caption{\label{fig:meas_equip} Experimental measurement setup schematic for (a) channel frequency response using a multiplied CW source and (b) channel efficiency using a chopped 77 and 300 K load.}
\end{figure}

As shown in Fig.~\ref{fig:meas_equip}a, a frequency-tunable CW source illuminates the hemispherical lens through the vacuum window to measure the frequency-response of each channel. A Hewlett-Packard (HP) 83752B synthesized sweeper generates CW signal between 0.01--20 GHz, which is then multiplied to 75--110 GHz by an HP 83557A mm-wave source. A Millitech rotary vane allows for variable attenuation of the output power of the rectangular WR-10 horn. A lock-in amplifier sends a reference signal to modulate the output of the sweeper and demodulates the resulting SQUID output of the detector system at that reference frequency. 

As shown in Fig.~\ref{fig:meas_equip}b, the channel efficiency is measured by alternating the source the detector views between two loads of known temperature. A 77~K liquid nitrogen bath is viewed through a 12.7 mm diameter aperture in a sheet of Eccosorb HR-10, and a rotating blade coated with Eccosorb HR-19 and held at room temperature periodically blocks that aperture, so the detector alternates between the 77 and 300~K loads at a fixed rate. The rotation is driven by a New Focus 3501 optical chopper assembly, which receives its reference signal from a lock-in amplifier; the same lock-in demodulates the resulting SQUID output.

\newpage % Forces LaTeX to break to the right column right here

\end{document}